\documentclass[12pt]{article}
\usepackage{scicite}
\usepackage{times}
\usepackage{graphicx}
\usepackage{amsmath}
\usepackage[usenames,dvipsnames]{color}
\usepackage[export]{adjustbox}

\newenvironment{sciabstract}{
\begin{quote} \bf}
{\end{quote}}

\newcounter{lastnote}

\newcommand{\bs}{\boldsymbol}

\title{Non-planar snake gaits: from S-starts to Sidewinding}

\author{N. Charles$^1$, R. Chelakkot$^2$,  M. Gazzola$^3$, B. Young$^4$, \& L. Mahadevan$^{5\ast}$
\\
\footnotesize{$^1 $ Engineering and Applied Sciences, Harvard University, Cambridge, MA 02138, USA}\\
\footnotesize{$^2$ Department of Physics, Indian Institute of Technology Bombay, Mumbai, 400076, India}\\
\footnotesize{$^3$ Department of Mechanical Science and Engineering, National Center for Supercomputing Applications, }\\
\footnotesize{Carl R. Woese Institute for Genomic Biology,  University of Illinois at Urbana-Champaign, }\\
\footnotesize{Urbana, IL, USA}\\
\footnotesize{$^4$ Department of Anatomy, A.T. Still University of Health Sciences,}\\
\footnotesize{ Kirksville College of Osteopathic Medicine, Kirksville, MO  63501, USA}\\
\footnotesize{$^5$ Engineering and Applied Sciences, Department of Physics, }\\
\footnotesize{Department of Organismic and Evolutionary Biology, Harvard University, Cambridge, MA 02138, USA}\\
\footnotesize{$^\ast$To whom correspondence should be addressed; E-mail:  lmahadev@g.harvard.edu}
}

\date{}
\begin{document}

\baselineskip16pt
\maketitle


\begin{sciabstract}
Of the vast variety of animal gaits, one of the most striking is the non-planar undulating motion of a sidewinder. But non-planar gaits are not limited to sidewinders. Here we report a new non-planar mode used as an escape strategy in juvenile anacondas ($Eunectes ~notaeus$). In the S-start, named for its eponymous shape, transient locomotion arises when the snake writhes and bends  out of the plane while rolling forward about its midsection without slippage. To quantify our observations, we present a  mathematical model for an active non-planar filament that interacts anisotropically with a frictional substrate and show that locomotion is due to a propagating localized pulse of a topological quantity, the link density.  A two-dimensional phase space characterized by scaled body weight and muscular torque shows that relatively light juveniles are capable of S-starts but heavy adults are not, consistent with our experiments. Finally, we show that a periodic sequence of S-starts naturally leads to a sidewinding gait. All together, our characterization of a novel escape strategy in snakes highlights the role of topology in locomotion, provides a phase diagram for mode feasibility as a function of body size, and suggests a role for the S-start in the evolution of sidewinding.
\end{sciabstract}

\newpage
Snakes exhibit a wide variety of gaits in a range of environments, of which the most commonly described ones are  rectilinear motion\cite{gray68,Hu2013}, periodic undulation \cite{gray68,Gray50,Gasc89,Moon98, Guo2008, biewener2018animal,Hu2009,alexander2013principles}, and  sidewinding\cite{gray68,Jayne86,Gans92,Marvi2014} on land. \textcolor{black}{More recently, there has been a recognition that this classification is incomplete \cite{jayne2020defines}, and needs to be broadened to include more complex modes of interaction such as arboreal climbing \cite{savidge2021lasso},  gliding in air \cite{yeaton2020undulation} etc. In addition, snakes employ a range of transient  modes such as concertina locomotion \cite{gray68,Jayne91} and various striking and escape lunges \cite{Alfaro2002,Alfaro2003,Irschick2005}, depending on the species, physical environment, and behavior.}

\begin{figure}
\includegraphics[width=\columnwidth]{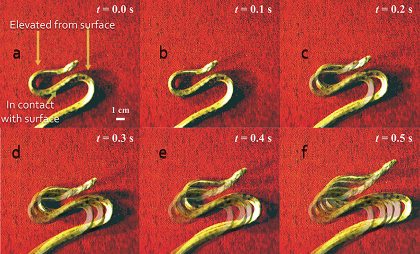}
\caption{\label{fig1}\linespread{1} \selectfont{}
{\footnotesize S-start of {newborn} anaconda ($Eunectes ~notaeus$) on a \textcolor{black}{red synthetic turf.} (a) The starting point of the locomotion is an `S' shape including three co-linear regions, connected by two curved regions that are elevated from the contact surface as the middle, straight segment pushes down against the surface. (b-f) Series of overlaid time-lapse images indicating the S-start (see Movie S1).} }
\end{figure}

Here we describe and quantify a new transient non-planar mode observed in newborn and juvenile yellow anacondas (\textit{Eunectes~notaeus}) that gives the snake a very fast forward velocity, often used defensively to escape a threatening situation. We dub this mode the \textit{S-start} since the snake forms an S-shape, by analogy to a transient maneuver commonly seen in juvenile fish, the C-start \cite{Domenici97,Gazzola2012}. To execute the S-start, \color{black}the snake forms an S-shape consisting of three roughly co-linear segments connected by two tightly curved regions as shown in Fig.~\ref{fig1}a. Motion begins when the two curved segments are lifted off the ground. Simultaneously the outer parallel straight segments move along the ground even as the middle straight segment remains stationary. This causes the curved segments to travel along the snake (see Movie S1). \textcolor{black}{Thus, the snake effectively moves  by propagating a spatially-localized region that is bent and twisted, similar to  a ruck in a rug \cite{Kolinski2009},  and propels itself forward (Fig.~\ref{fig1}b-f), as the S-shape flows along its length, until one end comes off the ground, terminating this transient mode of motility.} Globally, this leads to motion parallel to the direction of the snake's head and is reminiscent of sidewinding\cite{gray68,Marvi2014}, and as we will see, forms the building block from which many non-planar gaits including sidewinding may be constituted.
 

Lifting its body allows the snake to minimize frictional losses, but comes with a cost---the snake has to overcome gravity and move out of the plane to do so.  We expect slender snakes or those with relatively stronger musculature to be able to achieve this, while thicker snakes or those with less musculature/more bone to be unable to do so \cite{Calle2006}. \textcolor{black}{For a snake of mass density \( \rho \) and diameter \( a \), the maximum length \( \ell \) that can be supported out of the plane is determined by a balance between the gravitational torque \( \sim  \rho g a^2 \ell^2 \) and the active torque \( m \sim \Gamma a^2 h f(\phi)\), where \(\Gamma\) is the maximum muscular stress, \(\phi\) the fraction of tissue that is muscle, and \( h \sim a\) is the typical lift-off height, so that \(\ell \sim \sqrt{\Gamma a f(\phi)/(\rho g)}\). Therefore, as the snake grows in girth, its internal muscle fraction decreases (due to a concomitant increase in bone mass) \cite{Prange76,Anderson79,Garcia2006}, the length it can support off the ground decreases, and the S-start becomes physically unattainable. }
Indeed, our study of locomotion in three newborn, five juvenile, and two adult anacondas with a range of lengths and weights (SI) reveals that all the newborn and juvenile anacondas can move using the S-start, whereas none of the adult anacondas exhibit this type of locomotion.

 \begin{figure}
\begin{center}
\includegraphics[width=1.0\columnwidth]{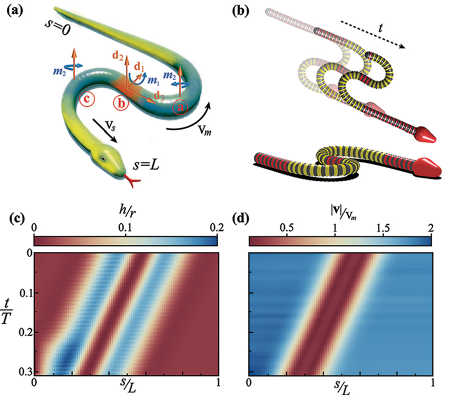}
\end{center}
\caption{\label{fig2}\linespread{1} \selectfont{}
{\footnotesize Mathematical model of the S-start motion and comparison with experiments. (a) A schematic showing localized in-plane torques at locations \(a\), \(c\) and a localized out-of-plane torque at \(b\), that propagate from head to tail synchronously at speed \(\text{v}_m\).  The velocity of snake center of mass \( \text{v}_s \) points parallel to head direction. (b) (top) Numerical simulations capture the S-start (see SI and Movie S2 for details). (bottom) Simulation snapshot shows the filament bending out of the plane. (c) Kymograph of the local vertical elevation as a function of arc length, \(s\), and time, \(t\), relative to characteristic time, \( T = L/\text{v}_m \), shows a minimum around \(b\) where the filament is stationary relative to ground, and is maximum at  \(a\) and \(c\). (d) Kymograph of the local speed with respect to the surface. The speed is nearly zero at region \(b\) and uniform for other segments.}}
\end{figure}

To understand these observations quantitatively, we model the snake (Fig.~\ref{fig2}a) as an active elastic filament  interacting frictionally with a planar substrate \cite{OReilly2017}. The filament is assumed to have a circular cross-section of diameter \(a\) and length \(L\) with aspect ratio \(L/a=50\). The axis is parametrized by a material coordinate \(s\), and the shape of the snake at a particular time \(t\) can be characterized in terms of a local position vector \(\textbf{r}(s,t)\) and an associated material orthonormal coordinate system \(\{\textbf{d}_1(s,t), \textbf{d}_2(s,t), \textbf{d}_3(s,t)\}\). 
We assume that the filament has a passive bending rigidity \(B\) and twisting rigidity \(C\), and we further neglect extensional and shear deformations of the filament, a reasonable assumption for long slender elastic objects, so that \(s\) becomes the arc length and \(\textbf{d}_3(s,t)\) is the tangent to the center-line of the snake.  
Since the filament is  capable of exerting muscular couples along its body, we define an active torque vector \(\textbf{m} = \Sigma_{i=1}^3\textbf{m}_i \textbf{d}_i\)  (Fig.~\ref{fig2}a), associated with the two bending modes and one twisting mode of the filament at each cross-section.  
Then, the motion of the filament is determined via the balance of linear and angular momentum governed by the equations
\(\rho a^2 \partial \textbf{v}(s,t)/\partial t=\partial \textbf{f}(s,t)/\partial s+ \textbf{F}_e(s,t)\) and \(\textbf{I} \partial \boldsymbol{\omega}(s,t)/\partial t= \partial \textbf{M}(s,t)/\partial s + \textbf{d}_3 \times \textbf{f}\). 
Here \(\rho\) is the density of the filament, \(\textbf{v} = \partial \boldsymbol{r}/\partial t\) is the local filament velocity, \(\textbf{f}\) is the vector of force resultants at a cross section, \(\boldsymbol{\omega}\) is the local angular velocity of filament rotation, \(\textbf{I}\) is the moment of inertia of the cross section (assumed to be circular), \(\textbf{M} = \textbf{B}\textbf{k}(s,t) + \textbf{m}(s,t)\) is sum of the passive elastic torques (associated with the vector of bending and twisting strains \(\textbf{k}(s,t)\) and the matrix of bending and twisting stiffnesses \(\textbf{B} = \text{diag}\left(B, B, C \right) \)) and the active muscular torques \( \textbf{m}(s,t) \) (SI), and \(\textbf{F}_e\) is the external force on the filament associated with gravity and frictional interactions with the ground (SI). 
We assume that the filament interacts with the substrate frictionally, with a finite friction coefficient that is anisotropic. Assuming \(\mu_f< \mu_b< \mu_s\) to be the coefficients of kinetic friction in the forward, backward and sideways directions respectively, we choose \(\mu_f:\mu_b:\mu_s=1.1:1.4:2\), based on  experiments \cite{Hu2009}.  

\begin{figure}
\begin{center}
\includegraphics[width=1.0\columnwidth]{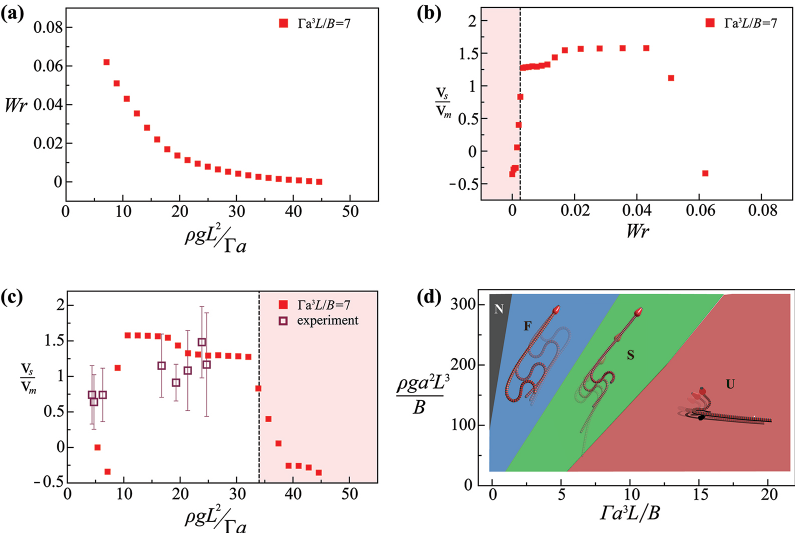}
\caption{\label{fig3} \linespread{1} \selectfont{} \footnotesize (a) Snake centerline \( Wr \) at one snapshot during simulated snake motion, as a function of its body weight relative to muscle torques, \(\rho g L^2/\Gamma a\).  As expected, \( Wr \) decreases monotonically with increasing weight and decreasing muscle.  (b) Forward speed of the snake relative to muscle speed, \(\text{v}_s/\text{v}_m\) as a function of body centerline \( Wr \).   \textcolor{black}{Below a certain \( Wr \) threshold, the snake fails to lift off the ground, and friction prohibits conversion of muscle propagation into forward motion; above a certain threshold, the snake flails, assuming a disordered three dimensional shape and failing to produce cohesive forward motion. For an intermediate range of \( Wr \) values, the snake lifts off the ground without flailing, maximizing speed.} (c) Forward snake speed as a function of body weight relative to muscle torque, combining (a) and (b).  Filled symbols are simulation data and open symbols correspond to yellow anaconda (\textit{Eunectes~notaeus}) assuming a value for \(\Gamma \text{ [kPa]} \simeq 7.72 (a\text{ [cm]})^{1.39}\) for snakes \cite{Moon2007}, measured on different surfaces, with the error bars of one standard deviation. The S-start fails in the shaded region, where the snake uncoils without a net translation and causes a backward shift of snake center-of-mass and a negative \(v_s\), and for large weight or small muscle torque.  Seen through the lens of parts (a) and (b), the S-start requires an intermediate value of \( Wr \), which is produced for intermediate weight to muscle ratios. (d) A two dimensional phase space indicates different locomotion regimes, characterized by two parameters, a scaled muscular torque \(\Gamma a^3 L/B\) and scaled weight \(\rho a^2 L^3/B\). Very small and very large values of the scaled weight are either inertially unstable (U) or so heavy that friction dominates (F), producing backwards motion.  For extremely heavy and weak snakes, i.e. \(\rho a^2 L^3/B \gg 1\), \(\Gamma a^3 L/B \sim  1\) no net motion is observed (N).  The S-start (S) is feasible only in an optimal regime of body weight and muscle forces (see Movie S2). Note that the vertical axis over the horizontal \((\rho a^2 L^3/B)/(\Gamma a^3 L/B)\) gives the horizontal axis in (c). That this single ratio determines the type of locomotion explains the straight phase boundaries in (d).}
\end{center}
\end{figure}

To complete the mathematical formulation of the problem, we note that the S-start shows the snake has two localized bends associated with the in-plane curvatures, \(\boldsymbol{m_2}\), at locations \((a)\) and \((c)\), and one localized bend associated with out-of-plane curvature at the contact zone in the central segment that is pressed downwards with an out-of-plane torque, \(\boldsymbol{m_1}\), at \((b)\) (Fig.~\ref{fig2}a). To model this arrangement of localized bending, we assume that the three active muscular torques are localized with a magnitude \(m \sim \Gamma a^3 e^{-\left(s-s_0-\text{v}_m t\right)^2/(2\sigma^2)}\), where \(\Gamma\) is the maximum magnitude of the muscular stress, \(s_0\) is the initial location of the torque, \(\sigma\) is the axial scale over which the torque acts.  We discretize the filament, implement friction, and solve the associated discretized governing equations using the frictional contact model and numerical integration scheme given in \cite{Gazzola2018}.

In Fig.~\ref{fig2}b, we show that these three localized torques move uniformly along the filament towards the tail with a speed \(\text{v}_m\) relative to the body, causing a pulse-like traveling wave that propels the filament parallel to itself with a speed \(\text{v}_s=\textbf{v}_{\text{cm}}\cdot \textbf{x}_h\) (\(\textbf{x}_h\) is the unit vector associated with the direction of the head \(s=0\); \(\textbf{v}_{\text{cm}}\) is the center-of-mass velocity of the snake) (see Movie S2).  
Eventually the active torques reach the tail and the whole filament straightens out.  In Fig.~\ref{fig2}c, we plot the local elevation of segments along the filament, showing that the traveling pulse corresponds to a small localized region of contact pushed down into the substrate while the curved regions are elevated above the surface.  In Fig.~\ref{fig2}d, we plot the local speed of the filament, \(|\textbf{v}|\), showing that the middle section at \((c)\) never slips relative to the surface, while the other regions move uniformly through this region. 

The localized pulse that underlies the S-start has a topological interpretation that stems from recognizing that the non-planar shape and motion of a filament can be characterized in terms of the CFW theorem \cite{Calugareanu1959} relating the topological link (\(Lk\)) to the twist (\(Tw\)) and writhe (\(Wr\)) via the formula \(Lk = Tw+Wr\). Here, we apply these quantities to our active filament by treating the filament centerline \( \textbf{r}(s, t) \) and director vector   \( \textbf{d}_1(s,t) \) as a mathematical ribbon (see SI for details).  In Fig.~\ref{fig3}a we show the \( Wr \) of the snake body at a single moment during the simulated motion; light, muscular snakes show much more \( Wr \) than heavier and less muscular snakes. In Fig.~\ref{fig3}b, we show that the scaled forward snake speed (averaged over a full motion cycle) is small for low values of \( Wr \) because of the effects of friction, and also small for large values of \( Wr \) because the snake tends to flail and lift off from the ground. However, the speed is maximized for an intermediate value of \( Wr \). \textcolor{black}{Combining the results of Fig.~\ref{fig3}a-b,  in Fig.~\ref{fig3}c we plot the scaled snake forward speed $\text{v}_s/\text{v}_m$ relative to muscular strength, and see that maximum forward speed is achieved for the weight and muscle that produce the optimal \( Wr \) in the snake body corresponding to a controlled three-dimensional snake body just short of flailing.}

To quantify the role of non-planar bending and writhing in the snake body during motion, we employ local densities of \(Lk\), \(Tw\) and \(Wr\), denoted by \(\lambda \), \( \tau\) and \(\omega\), respectively, and defined by 
\(Tw = \int_0^L \tau(s) ds,~ \lambda(s) = Lk(\bs{r}([s, s+\delta]))/\delta,~ \omega = Wr(\bs{r}([s, s+\delta]))/\delta\)
where \( \bs{r}([s, s+ \delta]) \) denotes the section of the snake centerline for which the arc length coordinate lies in the range \( [s, s + \delta] \), choosing \( \delta = L/10 \), determined empirically (see Fig.~SI-2 and Movie S2). In the S-start, \( \lambda(s,t) \neq 0 \) in a small zone, and as the snake moves forward, \( \lambda(s,t) \) translates along the body without changing its amplitude profile. In contrast, for friction-dominated locomotion \( \lambda_\text{friction}(s,t) \neq 0 \) and small over most of the snake body, while during unstable flailing, \( \lambda(s) \) shows no coherent pattern in space-time.  Together these observations show that a successful S-start has a simple topological interpretation: the snake must have a threshold, localized link current moving from its head to its tail. 

To characterize the feasibility of the S-start in terms of the snake's morphology and musculature, we characterize a phase space in terms of its scaled weight \(\rho  g a^2 \ell^3/B\) and scaled muscular torque \(\Gamma a^3\ell/ B\). In Fig.~\ref{fig3}d, we see that in a relatively light snake that is over-powered, inertia dominates over weight and friction, and the snake flails without much net movement   (Movie S2: part 3). For relatively heavy snakes that are under-powered, friction dominates inertia, and the active torques simply uncoil the snake into a straight shape, again leading to little net movement (Movie S2: part 2).  However, for intermediate weights, the S-start leads to a finite propulsion speed. These biomechanical limits of S-starts show how a juvenile can use S-starts for rapid escape, but lose that ability as they grow, as shown in Fig.~\ref{fig3}c. 

\textcolor{black}{The non-planar nature of the S-start has some similarities to a gait that leads to an illusion of the snake winding itself sideways, the eponymously named sidewinding gait. Here a snake lifts part of its body out of the plane, and places it askew periodically,   raising questions about the extent, origin and limits of this mode of locomotion from a dynamical, physiological and evolutionary perspective  \cite{gray68,Jayne86,Gans92,Alexader2003, biewener2018animal, Marvi2014,Tingle2020,jayne1988muscular,jayne2020defines}. Just as in the S-start, sidewinding involves spatially inhomogeneous non-planar transient motions. So could the S-start serve as a building block of a sidewinding gait? To probe this premise, we periodically initiate the spatiotemporal patterns of the torque triplets used for the S-start at the head and propagate them along the  body (Fig~SI-3a).  We find that as this wave propagates along the snake, it causes the body to translate orthogonally to its tail-to-head direction, pivoting about multiple contact points which are themselves stationary with respect to the ground (Fig.~SI-4, Movie S3).  We see that sidewinding involves multiple \( \lambda \)-pulses (Figs.~SI-2 and SI-4), each similar to an S-start \( \lambda\) profile propagating through the snake and indicative of how the sidewinding gait might have evolved. }

{\color{black} In this context, it is worth mentioning a very recent study~\cite{savidge2021lasso} that reports a novel arboreal mode of locomotion in climbing snakes,  facilitated by the propagation of strongly localized writhing and bending deformations. The similarity of this mode of locomotion with S-start in utilizing localized pulses as templates for transient movements further strengthens the case for the significance of $\lambda$ pulses on the evolution of three-dimensional gaits in snakes.}   

Our study highlights an unusual transient mode of locomotion in newborn and juvenile anacondas, emphasizing the role of out-of-plane motion in snake locomotion. A biophysical  model for non-planar modes allows us to quantify the mechanism by which this transient rapid motion can be achieved using a simple set of three localized propagating torques, and a phase diagram shows the regime where the S-start is feasible, consistent with observations. Our study also points to a pathway for the evolution of the sidewinding gait via periodic S-starts. More generally,  we see how developmental changes in animal size and shape lead to changes in the physical constraints that drive qualitative gait transitions. Understanding the biomechanical basis for the complex dynamics of contact of the snake with the substrate, the neuromechanical basis for the control of the S-start, and the potential ability to translate our observations to engineering solutions of rapid non-planar modes in active slender systems seem like natural next steps.

 {\bf Contributions.} B.Y. observed the novel behavior and carried out the experiments with snakes. L.M. conceived and designed the study and approaches.   N.C. performed the numerical simulations, in consultation with R.C., M.G. and L.M.  L.M. conceived of the mathematical model, the phase space and the scaling laws.  N.C. R.C. and L.M. wrote the paper. L.M. supervised the study. The authors declare no conflict of interest.

{\bf Acknowledgments.} We thank NSF grants  BioMatter DMR 1922321, MRSEC DMR 2011754 and EFRI 1830901, the Simons Foundation and the Henri Seydoux Fund (L.M.) for partial financial support. 

\end{document}